# Title: Cybersecurity of AI medical devices: risks, legislation, and challenges

Authors:
Elisabetta Biasin, KU Leuven Centre for IT & IP Law
Erik Kamenjašević, KU Leuven Centre for IT & IP Law
Kaspar Rosager Ludvigsen, University of Strathclyde

## Introduction

Medical devices and artificial intelligence (AI) systems rapidly transform healthcare provisions.[1] At the same time, due to their nature, AI in or as medical devices might get exposed to cyberattacks, leading to patient safety and security risks.[2] For example, an AI insulin pump under a cyberattack could stop working correctly and provoke serious health risks to the patient using it, which has been proven possible, as has defences against it.[3] Next to the health-related consequences that might turn fatal, cyberattacks on AI medical devices could also provoke indirect effects. These could range from diminishing patients' trust in the security of the healthcare system to hesitancy towards using these AI medical devices due to their cybersecurity-related vulnerabilities.[4]

The chapter is divided into three parts. The first part starts by setting the scene where we explain the role of cybersecurity in healthcare. Then, we briefly define what we refer to when we talk about AI that is considered a medical device by itself or supports one.[5] To illustrate the risks such medical devices pose, we provide three examples: the poisoning of datasets, social engineering, and data or source code extraction. In the second part, the paper provides an overview of the European Union

---

\* All the authors equally contributed to the writing process of this book chapter.

[1] For the ethical cybersecurity issues in healthcare, see Markus Christen, Bert Gordijn and Michele Loi (eds), *The Ethics of Cybersecurity*, vol 21 (Springer International Publishing 2020) <http://link.springer.com/10.1007/978-3-030-29053-5> , chapter 7.

[2] In this book chapter, we refer to 'AI as or in medical devices' following the caveats explained under section 'AI in or as medical devices'; see *infra;* Contrary to the EU, the United States' Food and Drug Administration (FDA) uses definitions of "Software as Medical Devices" as defined by the International Medical Device Regulators Forum (IMRDF), see International Medical Device Regulators Forum, '"Software as a Medical Device": Possible Framework for Risk Categorization and Corresponding Considerations' (2014). Definitions for when it is a medical device can be found in US Food and Drug Administration 'How to Determine if Your Product is a Medical Device' (*FDA.gov,* December 2019) <https://www.fda.gov/medical-devices/classify-your-medical-device/how-determine-if-your-product-medical-device>
accessed 11 August 2022. For critical analysis of this, see, e.g.,      Dhruv B Pai, 'Mapping the Genealogy of Medical Device Predicates in the United States' (2021) 16 PLOS ONE e0258153.

[3] Tamar Levy-Loboda and others, 'Personalized Insulin Dose Manipulation Attack and Its Detection Using Interval-Based Temporal Patterns and Machine Learning Algorithms' (2022) 132 Journal of Biomedical Informatics 104129.

[4] Elisabetta Biasin and Erik Kamenjašević, 'Cybersecurity of medical devices: new challenges arising from the AI Act and NIS 2 Directive proposals' [2022] International Cybersecurity Law Review, 163 <https://doi.org/10.1365/s43439-022-00054-x>.

[5] AI will be regulated separately, but it is worth noting that it can be considered a medical device or an accessory to one as well. For more, see Kaspar Ludvigsen, Shishir Nagaraja and Angela Daly, 'When Is Software a Medical Device? Understanding and Determining the "Intention" and Requirements for Software as a Medical Device in European Union Law' [2021] European Journal of Risk Regulation <https://doi.org/10.1016/j.clsr.2022.105656>.

[PRE-PRINT/PEER REVIEWED] 4.0 – 14/12/2022

(EU) regulatory framework relevant for ensuring the cybersecurity of AI as or in medical devices. Such a framework includes the Regulation (EU) 2017/745 on medical devices (MDR)[6], Directive (EU) 2016/1148 concerning measures for a high common level of security of network and information systems across the Union (NIS Directive)[7], Regulation (EU) 2019/881 on ENISA and on information and communications technology cybersecurity certification (Cybersecurity Act)[8], and Regulation (EU) 2016/679 on the protection of natural persons with regard to the processing of personal data and on the free movement of such data (GDPR)[9]. Two EU legislative proposals that, once adopted, will have an impact on the existing obligations concerning the cybersecurity of AI as or in medical devices are also addressed as part of such a framework. These are the EU AI Act proposal[10] and the NIS 2 Directive proposal[11]. Finally, the third part of the paper examines possible challenges stemming from the EU regulatory framework. In particular, we look toward the challenges deriving from the two legislative proposals and their interaction with the existing legislation concerning AI medical devices' cybersecurity. They are structured as answers to the following questions: (1) how will the AI Act interact with the MDR regarding the cybersecurity and safety requirements?; (2) how should we interpret incident notification requirements from the NIS 2 Directive proposal and MDR?; and (3) what are the consequences of the evolving term of critical infrastructures?[12]

## Cybersecurity in Healthcare

As the uptake of medical devices that are part of the Internet of Things (IoT) or otherwise include hardware and software increases, so does the potential and reality of the cybersecurity landscape which lies directly behind it.[13] In cybersecurity, it is well established that anything with an operating system and perhaps network access is vulnerable, which then includes anything with these attributes which are used in healthcare.[14] Like security, cybersecurity involves adversaries and defenders, and the risks presented by the failures to mitigate or prevent the consequences from arising can, in healthcare, also affect the physical and mental health of patients. One future noteworthy challenge is the potential use of quantum computers, which will make a majority of past encryption and security schemes obsolete by the exponential increase in computational power, but new quantum-proof defences may themselves be vulnerable to conventional cyberattacks on top.[15] Cybersecurity is, in this sense, both a necessity, to prevent the most trivial attacks and measures, a financial burden on manufacturers and healthcare authorities, and merit, which is regulated by medical device legislation.[16] We can only know about the weaknesses and potential new attack venues if these are rigorously tested, which is why cybersecurity research is essential to assess and potentially prevent future disasters. Just because something has not been discovered in practice yet, does not mean it can or has not already happened.

---

[6] Regulation (EU) 2017/745 of the European Parliament and of the Council of 5 April 2017, on medical devices, amending Directive 2001/83/EC, Regulation (EC) No 178/2002 and Regulation (EC) No 1223/2009 and repealing Council Directives 90/385/EEC and 93/42/EEC [2017] OJ L 117/1; A soft law document relevant for the interpretation of different provisions, roles and obligations in the context of the medical devices' cybersecurity is the Medical Device Coordination Group Guidance on cybersecurity for medical devices MDCG 2019-16, Rev.1, Guidance on Cybersecurity for medical devices, July 2020 (Medical Device Coordination Group, 'MDCG 2019-16 Guidance on Cybersecurity for Medical Devices, December 2019' (2020), hereinafter 'MDCG guidance'). The MDCG Guidance deals with the cybersecurity-related provisions embedded in the MDR. This non-binding document provides a comprehensive overview of cybersecurity-related requirements that manufacturers must implement to comply with the MDR and to ensure the appropriate level of the cyber resilience of the medical device.

[7] Directive (EU) 2016/1148 of the European Parliament and of the Council of 6 July 2016, concerning measures for a high common level of security of network and information systems across the Union [2016] OJ L 194/1.

[8] Regulation (EU) 2019/881 of the European Parliament and of the Council of 17 April 2019 on ENISA (the European Union Agency for Cybersecurity) and on information and communications technology cybersecurity certification and repealing Regulation (EU) No 526/2013 (Cybersecurity Act) [2019] OJ L 151/15.

[9] Regulation (EU) 2016/679 of the European Parliament and of the Council of 27 April 2016 on the protection of natural persons with regard to the processing of personal data and on the free movement of such data, and repealing Directive 95/46/EC (General Data Protection Regulation) [2016] OJ L 1191.

[10] Commission, 'Proposal for a Regulation of the European Parliament and of the Council Laying Down Harmonised Rules on Artificial Intelligence (Artificial Intelligence Act) and Amending Certain Union Legislative Acts' COM/2021/206 final.

[11] Commission 'Proposal for a Directive of the European Parliament and of the Council on measures for a high common level of cybersecurity across the Union, repealing Directive (EU) 2016/1148' COM(2020) 823 final.

[12] This book chapter lists three legal challenges. There could be others insisting on medical device cybersecurity, such as cybersecurity certification, see Elisabetta Biasin and Erik Kamenjašević, 'Cybersecurity of Medical Devices: Regulatory Challenges in the European Union' in Carmel Shachar and others (eds), The Future of Medical Device Regulation: Innovation and Protection (Cambridge University Press 2022) <https://www.cambridge.org/core/books/future-of-medical-device-regulation/cybersecurity-of-medical-devices/AC01289C2DB05E44D0D98A9E66666562>. Therefore, the list should be regarded as non-exhaustive.

## AI in or as Medical Devices

AI can be defined in a plethora of ways, and with the event of the upcoming EU AI Act, actual legal definitions are proposed.[17] This is fascinating, as some of the AI which are ML or purely statistics-based will not *per se* be considered AI by those who produce them.[18] Regulating software or code that can be considered everything from classical decision-making systems to very simple statistical analysis may be a competitive advantage for the EU, as it allows for tighter regulation than the general rules that apply, e.g., software.[19] Due to the network effects of strong European regulation, the AI Act proposal would then set the standard for AI globally.[20] But there is also a risk that the AI Act proposal only considers certain types of systems, such as those that purely work through ML.[21] This would leave the rest out and imply that these be regulated by other product legislation, such as the MDR. Either direction will have direct consequences for AI as or in medical devices, as the manufacturers await either more regulation with the current approach or less regulation if Annex I is severely limited. Coincidentally, it would also have an impact on a cybersecurity level, as the AI Act proposal does have a literal cybersecurity reference in Article 15, while other types of product legislation may not. It seems likely that the possibility of tighter regulation would lead to less risk for the patient or others on which the AI is used, despite the lack of the patient being a purpose in the MDR or individuals in the proposed AI Act, because of the difference between regulating cybersecurity literally and delegating it to guidance.[22]

---

[13] On the uptake of medical devices, IoT and AI, see eg, Roman V Yampolskiy, 'AI Is the Future of Cybersecurity, for Better and for Worse' [2017] Harvard Business Review <https://hbr.org/2017/05/ai-is-the-future-of-cybersecurity-for-better-and-for-worse> accessed 25 October 2022.

[14] Ross Anderson, Security Engineering: A Guide to Building Dependable Distributed Systems (John Wiley & Sons 2020), 15.

[15] On security and quantum computers see Celi S, 'The Quantum Solace and Spectre' (The Cloudflare Blog, 21 February 2022) <http://blog.cloudflare.com/quantum-solace-and-spectre/> accessed 25 October 2022 (describing how companies like Cloudflare have already started developing relatively future proof techniques against it; Wouter Castryck and Thomas Decru, 'AN EFFICIENT KEY RECOVERY ATTACK ON SIDH (PRELIMINARY VERSION)' <https://eprint.iacr.org/2022/975.pdf> accessed 25 October 2022.

[16] Usually only in the form of guidance, although this may be changing in a EU context at least.

[17] Pursuant to Article 3(1) and Annex I of the AI Act proposal, we can define AI as either models based on Machine Learning (a), logical or knowledge-based systems (b) or statistical models (c), as has been noted in other chapters of this book.

[18] see Peter Savadjiev and others, 'Demystification of AI-Driven Medical Image Interpretation: Past, Present and Future' (2019) 29 European Radiology 1616, 1616–1618.

[19] In lieu of the "Brussels effect", as AI will be developed with EU sales in mind on a worldwide basis. See eg, Lee A Bygrave, 'The "Strasbourg Effect" on Data Protection in Light of the "Brussels Effect": Logic, Mechanics and Prospects' (2021) 40 Computer Law & Security Review 105460 for more. Furthermore, analogous to the discussions which the GDPR fostered concerning data analysis in general, and the network effects which it has had going forward, see eg Christian Peukert and others, 'Regulatory Spillovers and Data Governance: Evidence from the GDPR' (2022) Marketing Science <https://pubsonline.informs.org/doi/pdf/10.1287/mksc.2021.1339> accessed 25 October 2022.

[20] see, eg, Michal S Gal and Oshrit Aviv, 'The Competitive Effects of the GDPR' (2020) 16 Journal of Competition Law and Economics 349.

The damage which breaches in cybersecurity on AI can have should dictate how the lifecycle of the systems are regulated. AI as a medical device make for an interesting case because of how they always interface with patients and users. An analogy can explain this further. AI that are part of or steer industrial control systems or cyber-physical systems may be regulated via the AI Act in a similar way to AI in or as medical devices. But there is a crucial difference in AI that control such systems do not always put humans at risk in such a bodily invasive manner, as some medical devices like surgical robots and implantable medical devices do.[23] What makes AI as or in medical devices unique is the kind of *risks* they pose, as the models and methods used to create and control them are not unique or novel. Examples of these physical risks could be physical harm to the patient through injuries, wrong or dangerous recommended medication, and inadequate choice of surgery.[24]    These failures in cybersecurity, make the usage of such medical devices more dangerous than if they possessed no hardware or software, and AI as medical devices go further than this. Instead of just attacking a local network or device, the attacker can cause damage to the model, or the AI service used by or as the medical devices, causing damage at scale instead of locally.

The Medical Device Coordination Group (MDCG) has made an overview of some extremely serious but likely common types of cyberattacks and their consequences on patients in Annex II of their MDCG Guidance on Cybersecurity.[25] Many of these will apply directly to AI as or in medical devices. Next to these, privacy risks could include leakage of special categories of personal data or profiles necessary for ML which the AI has created and specific information about movement or whereabouts.[26] Physical

---

[21] Indicated by position papers such as ECOMMERCE EUROPE, 'Position Paper on the Artificial Intelligence Act' (2022) <https://ecommerce-europe.eu/wp-content/uploads/2022/02/ECOM-AI-Position-Paper-01022022.pdf> accessed 25 October 2022.

[22] see AI Act proposal, art 1.

[23] They, nevertheless, bring risks to individuals: see, eg, Kazukuni Kobara, 'Cyber Physical Security for Industrial Control Systems and IoT' (2016) E99D IEICE Transactions on Information and Systems 787.

[24] Assuming AI will assist with telerobotic surgery, eg, Homa Alemzadeh and others, 'Targeted Attacks on Teleoperated Surgical Robots: Dynamic Model-Based Detection and Mitigation' (2016) Proceedings - 46th Annual IEEE/IFIP International Conference on Dependable Systems and Networks, DSN 2016 395. See also eg, Sujata Swain and Rajdeep Niyogi, 'SmartMedicist: A Context-Aware System for Recommending an Alternative Medicine' (2018) 14 International Journal of Pervasive Computing and Communications 147. The consequences of such a system suffering various failures can be detrimental.

[25] see MDCG guidance (n 6). For more information on the effect of the guidance, see Kaspar Rosager Ludvigsen and Shishir Nagaraja, 'Dissecting Liabilities in Adversarial Surgical Robot Failures: A National (Danish) and EU Law Perspective' (2022) 44 Computer Law and Security Review <https://doi.org/10.1016/j.clsr.2022.105656>.

[26] On privacy risks including leakage of special categories of personal data, see Hang Xu and others, 'Transferable Environment Poisoning: Training-Time Attack on Reinforcement Learning' (2021) 3 Proceedings of the International Joint Conference on Autonomous Agents and Multiagent Systems, AAMAS 1386; on leakage of profiles necessary for ML, see Xinlei

risks can be caused by indirect consequences, such as the wrong calibration of instruments of a surgical robot, the same for medicine recommender systems, or other types of communication disruption which may end up causing indirect injuries.[27] Essentially, normal accidents or non-adversarial risks exist in the exact same ways for AI as or in medical devices as they do for medical devices without them, but AI has an additional layer of complexity on top in the form of its coding and potentially ML and because of its digital presence in the form of its cybersecurity.[28]

## Specific Cybersecurity Issues

AI, in general, presents specific risks within cybersecurity. One which we will not go further into is causing the AI to malfunction or otherwise make a wrong decision. In the ML context, this is often caused by an "adversarial sample/examples", popularised through road signs that a car vision would misread or causing a system to misidentify pictures.[29] These purely cause the AI to fail, and while these are theoretically common, they do not represent the complexity that our new three possess.

The three adversarial cyberattacks we use as examples are not exhaustive and present overarching categories in themselves.[30] Poisoning is known as the biggest threat to ML-based systems, and if many types of AI used in healthcare make use of such decision-making mechanisms, considering this specific vulnerability is paramount. Social Engineering is the most common type and covers everything from shoulder-surfing to phishing, attacks which everyone should know about and whose defences involve humans and behavioural and organizational training. Finally, extraction is another unique ML vulnerability, which leaves both the source code and training data vulnerable, the latter may cause both practical and data protection issues if leaked.

## Poisoning of Datasets

If the AI, which is used as a medical device or together with one, uses ML, it will use datasets to create its classifiers which it then uses to make decisions from. This can be made further complicated with neural networks, but what matters is the distinction between whether it uses such datasets or not. If

---

Pan and others, 'How You Act Tells a Lot: Privacy-Leaking Attack on Deep Reinforcement Learning' (2019) 1 Proceedings of the International Joint Conference on Autonomous Agents and Multiagent Systems, AAMAS 368; leakage of information about movement and whereabouts is increasingly becoming a danger, and it would give the same risks as general stalkerware, see Cynthia Khoo, Kate Robertson and Ronald Deibert, 'Installing Fear: A Canadian Legal and Policy Analysis of Using, Developing, and Selling Smartphone Spyware and Stalkerware Applications' (2019) <https://citizenlab.ca/2019/06/installing-fear-a-canadian-legal-and-policy-analysis-of-using-developing-and-selling-smartphone-spyware-and-stalkerware-applications/> accessed    25 October 2022.

[27] For a discussion on how to prove and understand injuries caused by adversarial cyberattacks, indirectly or directly, see Ludvigsen and Nagaraja (n 27).

[28] From a terminological point of view, 'adversarial' can be translated to opponent, and refers to actors who want to attack a given software or hardware system. 'Non-adversarial' refers to failures which occur without them.

[29] On adversarial examples, see Richard Tomsett and others, 'Why the Failure? How Adversarial Examples Can Provide Insights for Interpretable Machine Learning' [2018] 2018 21st International Conference on Information Fusion, FUSION 2018 838; see also Dawei Zhou and others, 'Removing Adversarial Noise in Class Activation Feature Space', *2021 IEEE/CVF International Conference on Computer Vision (ICCV)* (IEEE 2021) <https://ieeexplore.ieee.org/document/9710484/> accessed 25 October 2022.

[30] Other authors use four categories, but these do not reflect the most common types of attacks, as these are always present, see Anderson (n 14) 876–877.

it does, it is vulnerable to adversarial cyberattacks, which can poison these datasets.[31] This is usually in the form of manipulation, where the data is changed, which will be reflected in the decisions the AI later takes but can take other forms such as inserting random numbers or deleting parts of it instead.[32]

Constructing and mounting generic machine-learning based attacks related to medical devices was shown as far back as 2015, where Mozaffari-Kermani et al. experimentally showed that generic poisoning, in the form of purely adding to datasets, either derived or known, are possible on a wide range of systems.[33] Most interestingly, the experiments were done on datasets used in healthcare settings, which both currently, but especially in the future, will include machine-learning based systems in various forms. The attack consists of two algorithms, the first describes the insertion, while the second dictates how the malicious insertions are designed. As with all machine-learning attacks, deception is key, and the effects of the attack should be shown in the outcome. In this case, the accuracy of the machine-learning models decreases by as much as 24% and should be realistic even in a real-world setting.

In the best of situations, no injury is caused by the attack and the AI merely functions wrongly or does nothing. In the worst, it could lead to injury or danger to the patient. This can be done in very subtle ways that may not be detectable or in an obvious manner that will by itself make the AI malfunction or otherwise fail.[34] Currently, any software which makes use of learning or datasets in general to function will be vulnerable to this attack.[35] Unlike simpler systems, AI that uses learning is wholly reliant on this data being usable and worthwhile unless a human fully controls it. However, at that point, it may not qualify to be considered AI in the first place.[36] There are no known examples of this attack being successful, but it is widely regarded as the most significant vulnerability to ML-based systems.[37]

## Social Engineering

Social engineering is the art of either attacking a system purely with human attributes, or the first step in a series of attacks which make use of other techniques after one has penetrated the defences. Ontologically, social engineering attacks has 6 factors, in no particular order: *Target, compliance*

---

[31] For a general overview, see Matthew Jagielski and others, 'Manipulating Machine Learning: Poisoning Attacks and Countermeasures for Regression Learning' (2018) 2018-May Proceedings - IEEE Symposium on Security and Privacy 19.

[32] Make no mistake, this is a very diverse and a complicated area of cybersecurity. Like elsewhere, there are several subdivisions of different cyberattacks which can be done to different learning models.

[33] Mehran Mozaffari-Kermani and others, 'Systematic Poisoning Attacks on and Defenses for Machine Learning in Healthcare' (2015) 19 IEEE Journal of Biomedical and Health Informatics 1893.

[34] see Efrat Shimron and others, 'Subtle Data Crimes: Naively Training Machine Learning Algorithms Could Lead to Overly-Optimistic Results' (2021) <http://arxiv.org/abs/2109.08237> accessed     25 October 2022. For examples, see in Sara Kaviani, Ki Jin Han and Insoo Sohn, 'Adversarial Attacks and Defences on AI in Medical Imaging Informatics: A Survey' (2022) Expert Systems With Applications 116815 <https://doi.org/10.1016/j.eswa.2022.116815>.

[35] And this is a feature of anything which makes use of these techniques, and there are no means to generally prevent it, see Andrew Ilyas and others, 'Adversarial Examples Are Not Bugs, They Are Features', *33rd Conference on Neural Information Processing Systems* (2019).

[36] This is not to be confused with `Supervised Learning', which will still enable the AI to fulfil the legal definition in Annex I of the AI Act proposal, regardless of whether it is changed drastically.

[37] Anderson (n 14), 875 - 879

[PRE-PRINT/PEER REVIEWED] 4.0 – 14/12/2022

*principles, techniques, goal, medium and social engineer*.[38] Target is individual or organisation, and compliance principles refer to what you are trying to abuse, be it friendship, internal order, or rules. Techniques concern anything from phishing to baiting or directly cheating individuals through conversation, while goal is either financial, access or disruption. Medium is anything from shoulder-surfing, online, phone or snail mail, while social engineer concerns whether the attack is done by an individual or a group.

Social engineering is a general threat to most systems, either through phishing or outright guessing passwords and means to access systems out of the relevant individuals.[39] The focus has in recent years been on general, or spear phishing, which has yet to subside or reduce in the currently massive numbers, where the accounts of senior staff or others are mimicked, and the victim is convinced to either click links, transfer funds or documents to the adversary.[40] The technique that enables extremely trustworthy looking emails or messages is sometimes referred to spoofing, as the receiver must believe that the email is both from the right server and the right domain.[41] They do so to identify what is required to commit to the spear-fishing attack, in this case whether the targeted mail server could tell the real domain from a fake with one less character, which it failed to do.

There has been a rather large number of successful social engineering attacks on healthcare providers and companies caused by phishing, but one of particular note is the one that successfully hit Magellan Incorporated in 2020.[42] The number of affected individuals was significant, and the attackers had the opportunity not only to deploy ransomware but could have also taken control of any deployed software such as medical devices run or supported by AI, or even poisoned the datasets or manipulated the machine-learning models. There has been no reporting indicating this was the case, but it remains as one of the of the more serious breaches, due to the unique services which Magellan offer, either through IoT or CPS, the latter specifically surgical robots. The number of individuals who can be affected by privacy breaches only grows, but so does the potential that these rather simple attacks have when more sophisticated or widespread technology is deployed by these same entities.[43] What makes it vital here is the risks that access to such AI would give through these attacks. An adversary through social engineering could get access to the system with the intent to, for example, harm a specific patient or a group. These attacks can be harmful enough via financial or reputational losses, but unless the AI has no control over what harm humans can directly or indirectly suffer, the risk from these types of attacks persists. Mitigation exists, and they mostly rely on the very humans in the loop within the system where the AI resides.[44] These attacks are often used to either escalate and gain control of the system or perhaps enable the other two examples of attack in this text.[45]

---

[38] This specific framework and understanding comes from figure 2 in Francois Mouton and others, 'Social Engineering Attack Framework', *2014 Information Security for South Africa* (IEEE 2014) <http://ieeexplore.ieee.org/document/6950510/> .

[39] Jan-Willem Bullée and Marianne Junger, 'Social Engineering' in Thomas J Holt and Adam M Bossler (eds), The Palgrave Handbook of International Cybercrime and Cyberdeviance (Springer International Publishing 2020) <http://link.springer.com/10.1007/978-3-319-78440-3_38> accessed 25 October 2022.

[40] Concerning spear phishing, there exists a myriad of good human behaviour focused literature, as well as technical, see eg Yuosuf Al-Hamar and others, 'Enterprise Credential Spear-Phishing Attack Detection' (2021) 94 Computers & Electrical Engineering 107363. See also Hossein Abroshan and others, 'COVID-19 and Phishing: Effects of Human Emotions, Behavior, and Demographics on the Success of Phishing Attempts During the Pandemic' (2021) 9 IEEE Access 121916.

[41] Al-Hamar and others (n 42).

[42] see HIPAA Journal, 'Healthcare Data Breaches Due to Phishing' (HIPAA Journal, nd) <https://www.hipaajournal.com/healthcare-data-breaches-due-to-phishing /> accessed 25 October 2022; Sergiu Galtan, 'Healthcare giant Magellan Health hit by ransomware attack' (Bleeping Computer, 12 May 2020) <https://www.bleepingcomputer.com/news/security/healthcare-giant-magellan-health-hit-by-ransomware-attack/>, accessed 25 October 2022.

[43] Successful adversarial attacks like these focus on leeching information and reselling it at a later point.

### Extraction of Data or Source Code

AI as or in medical devices make decisions. This is a core reason to use them, and it is one of the advantages they have over non-deciding systems. But with this comes a unique weakness, which is that there exists a myriad of adversarial attacks which can lure out the classifiers or the very core of the AI.[46] In practice, the adversary simply asks or tricks the AI into revealing this or amount to faithful and useful reconstructions of the model. These attacks can be black-box (no knowledge of the model) or white-box (partial knowledge of the model), illustrated by the two classic papers by Tramer et al.[47] In the black-box attack, analysis of what information the providers of machine-learning provide is used to predict what the model does with the data through queries, and in particular through what the queries indirectly or inversely say about the model.[48] This is then used to reconstruct a model which can provide the same functionality as the models which the attacker does not know, through depictions of the decision tree, leaves of the decision tree, categories, labels and other relevant information. The white-box paper goes further, and while it requires knowledge of the models (though not full), it discusses the idea of transferable extraction or otherwise adversarial example attacks, to cause the same consequences as the past paper, but on a myriad of different models.[49] Surprisingly, it finds proof of transferability, but also the opposite, that there are limits to how easy attacks that work on one type can be transferred to others. Through the extraction of either data or the algorithms, there is, at the very minimum, a privacy risk or other potential physical threats, as the data could contain information about the patient.[50] The other special risk comes from the re-building of the model, as the adversary can use the same techniques to recreate the targeted AI and either uses this information to find other vulnerabilities or simply sell it to a competitor of the manufacturer.[51] Unlike social engineering attacks, there are no major mitigation measures, and they will always exist if the AI is built to respond to inputs.[52] Indeed, it is through these very inputs that the attacks happen, and it is a systemic vulnerability of any ML-based system. Finally, there is also a unique immaterial risk, as competitors of manufacturers of the AI could use these same attacks to give themselves a competitive advantage, making the type of adversaries more diversified than in the other examples. There are no known examples of these attacks in practice, but they remain a serious threat for the reasons above.

---

[44] For an overview in general, which will apply to AI directly, Walter Fuertes and others, 'Impact of Social Engineering Attacks: A Literature Review' in Álvaro Rocha, Carlos Hernan Fajardo-Toro and José María Riola Rodríguez (eds), Developments and Advances in Defense and Security (Springer Singapore 2022).

[45] Kang Leng Chiew, Kelvin Sheng Chek Yong and Choon Lin Tan, 'A Survey of Phishing Attacks: Their Types, Vectors and Technical Approaches' (2018) 106 Expert Systems with Applications 1.

[46] Nicolas Papernot and others, 'Practical Black-Box Attacks against Machine Learning' [2017] ASIA CCS 2017 - Proceedings of the 2017 ACM Asia Conference on Computer and Communications Security 506.

[47] Florian Tramer and others, 'Stealing Machine Learning Models via Prediction APIs', *Proceedings of the 25th USENIX Security Symposium* (2016); Florian Tramer and others, 'The Space of Transferable Adversarial Examples' (arXiv, 23 May 2017) <http://arxiv.org/abs/1704.03453>.

[48] Tramer and others, 'Stealing Machine Learning Models via Prediction APIs' (n 49).

[49] Tramer and others, 'The Space of Transferable Adversarial Examples' (n 49).

[50] Zainab Hussein Arif and others, 'Comprehensive Review of Machine Learning (ML) in Image Defogging: Taxonomy of Concepts, Scenes, Feature Extraction, and Classification Techniques' (2022) 16 IET Image Processing 289.

[51] Xueluan Gong and others, 'Model Extraction Attacks and Defenses on Cloud-Based Machine Learning Models' (2020) 58 IEEE Communications Magazine 83.

[52] Alice Hutchings, Sergio Pastrana and Richard Clayton, 'Displacing Big Data: How Criminals Cheat the System', The Human Factor of Cybercrime (Routledge 2017).

### The EU Legal Framework Relevant for the Cybersecurity of AI in or as Medical Devices

The EU legal framework having provisions that are of relevance to the cybersecurity of medical devices consists of the Regulation (EU) 2017/745 on medical devices (MDR)[53], Directive (EU) 2016/1148 concerning measures for a high common level of security of network and information systems across the Union (NIS Directive)[54], Regulation (EU) 2019/881 on ENISA and on information and communications technology cybersecurity certification (Cybersecurity Act)[55], and Regulation (EU) 2016/679 on the protection of natural persons with regard to the processing of personal data and on the free movement of such data (GDPR)[56]. Next to these already applicable laws, the two legislative proposals that the EU legislator currently discusses are likely to apply to AI medical devices: the EU AI Act proposal[57] and the NIS 2 Directive proposal[58]. All of them are briefly elaborated on below by presenting their most relevant provisions concerning the cybersecurity of medical devices.

*Current laws*

The Cybersecurity Act aims to ensure the proper functioning of the EU internal market by reaching a high level of cybersecurity of network and information systems, communication networks, services, and devices within the Union. One of the Cybersecurity Act's objectives is to create a new framework for European Cybersecurity Certificates of ICT products, processes, and services. Certification schemes established under the Act's framework are voluntary, and vendors can decide whether they would like their products to be certified. The Act mainly impacts the AI medical devices' manufacturers since these devices may fall under the definition of an ICT product, being an 'element of a network of information systems', which consequently implies the application of the NIS Directive.[59] Healthcare providers may also fall within the scope of CSA since they use ICT processes or ICT services to carry out their activities.[60] To obtain cybersecurity certification, manufacturers or healthcare providers may voluntarily (when not prescribed by national or EU law) apply to the conformity assessment bodies of their choice established in the EU. Such a body produces a formal evaluation of the device against a defined set of criteria and standards, which results in issuing a certificate that should help assure users and maintain the trust and security of the device.

The NIS Directive is relevant to cybersecurity considerations concerning network and information systems. Specifically, the Directive lays down obligations for the EU Member States to reach a minimum harmonisation level across the EU concerning network and information systems' security. Network and information systems are defined as 'any device or group of interconnected or related devices, one or more of which, pursuant to a program, perform automatic processing of digital data.'[61] Applied to the healthcare sector, this definition of network and information systems may encompass hospitals' IT networks as well as AI medical devices. The NIS Directive sets security and notification requirements for operators of essential services (OES). Healthcare providers are concerned by the NIS Directive when identified as OES by the respective Member State. As OES, healthcare providers have to ensure a minimum level of security for their networks and information systems and have to notify security incidents to competent authorities without delay. To reach that level, they must take appropriate and proportionate technical and organizational measures to manage the risk posed to the NIS security, which they use in their operations. Security measures should be taken with regard to the state of the art, and their purpose is to prevent and minimize the impact of incidents affecting the security of the NIS used for the provision of essential services to ensure the continuity of those services.[62] These measures and modalities for communication of security incidents are defined at a national level by each Member State, which is required to adopt national strategies on network and information security.[63]

---

[53] see n 7.

[54] see n 9.

[55] See n 10.

[56] see n 11.

[57] see n 12.

[58] see n 13.

[59] CSA, art 2(12).

[60] Furthermore, see CSA art 56(3) which sets healthcare as a focus priority for the European Commission.

[61] NISD, art 4.

The MDR sets general requirements that may implicate cybersecurity considerations. They primarily address manufacturers of medical devices.[64] Article 5(1) MDR obliges manufacturers to ensure that the device complies with the MDR obligations when used following its intended purpose. According to Article 5(2) MDR, a medical device shall meet the general safety and performance requirements set out in Annex I MDR, taking into account the intended purpose.[65] As part of the general requirements set in Annex I MDR, 'devices shall achieve the performance intended by the manufacturer' and be designed in a way suitable for the intended use.[66] They shall be safe and effective, and associated risks shall be acceptable when weighed against the benefits of the patients and level of protection of health and safety while taking into account the state of the art.[67] Moreover, 'manufacturers shall establish, implement, document, and maintain a risk management system'.[68] Part of this system also includes risk control measures to be adopted by manufacturers for the design and manufacture of a device, and they shall conform to safety principles and state of the art.[69] A medical device designed to be used with other devices/equipment as a whole (including the connection system between them) has to be safe and should not impair the specified performance of the device.[70] Furthermore, a medical device shall be designed and manufactured to remove, as far as possible, risks associated with possible negative interaction between software and the IT environment within which they operate.[71] If a medical device is intended to be used with another device, it shall be designed, so the interoperability and compatibility are reliable and safe.[72] A medical device incorporating electronic programmable systems, including software or standalone software as a medical device, 'shall be designed to ensure repeatability, reliability, and performance according to the intended use'.[73] '[A]ppropriate means have to be adopted to reduce risks or impairment of the performance'.[74] A medical device should be developed and manufactured according to the state of the art and by respecting the development life cycle principles, risk management (including information security), verification, and validation.[75] Finally, manufacturers must 'set out minimum requirements concerning hardware, IT network characteristics, and IT security measures, including protection against unauthorized access'.[76] Concerning information to be supplied together with the device, manufacturers must inform about residual risks, provide warnings requiring immediate attention on the label and, for electronic programmable system devices, give information about minimum requirements concerning hardware, IT networks' characteristics and IT security measures (including protection against unauthorized access), necessary to run the software as intended.[77]

---

[62] see, NIS Cooperation Group, 'Reference document on security measures for Operators of   Essential Services' (2018). Concerning the term 'state of the art': the term defines that the technology used must be the best possible within a certain context, which depends on the market, type of technology and threats. For more, see Sandra Schmitz, 'Conceptualising the Legal Notion of "State of the Art" in the Context of IT Security' in Michael Friedewald and others (eds), *Privacy and Identity Management. Between Data Protection and Security*, vol 644 (Springer International Publishing 2022). <https://link.springer.com/10.1007/978-3-030-99100-5_3> accessed 25 October 2022. The term also received some unwarranted criticism from the privacy community in a GDPR context, see Kutyłowski, Mirosław and others, `GDPR – challenges for reconciling legal rules with technical reality', European Symposium on Research in Computer Security (Springer, Cham, 2020), which sadly fails to consider the historical connotation state of the art as a term has.

[63] NISD, art 1. A national strategy is according to the NISD a 'framework providing strategic objectives and priorities on the security of network and information systems at national level' (NISD, art 4).

[64] MDR, art 2(30). A manufacturer is 'the natural or legal person who manufactures or fully refurbishes a device or has a device designed, manufactured, or fully refurbished and markets that device under its name or trademark'.

[65] MDR, art 5(2).

[66] MDR, Annex I, req 1.

[67] ibid. The intended purpose is defined in Article 2(12) MDR as 'the use for which a device is intended according to the data supplied by the manufacturer on the label, in the instructions for use or in promotional or sales materials or statements and as specified by the manufacturer in the clinical evaluation'.

[68] ibid, req 3.

[69] ibid, req 4.

[70] ibid*.,* req 14.1.

The GDPR is relevant to AI medical devices' cybersecurity because these devices function by processing a vast amount of personal (and non-personal) data. In such a case, the Regulation lays down rules for protecting natural persons regarding the processing of their personal and sensitive data, including data concerning health. Amongst the many requirements, the GDPR sets obligations to ensure the security of the processing of data, which takes place in the lifecycle of AI medical devices.[78] Parties involved in the processing of personal data must put in place technical and organizational measures that are adequate to the risk of the processing. Security measures and obligations include the performance of risk assessments and the notification of a personal data breach to competent authorities. In the case of AI medical devices, security measures may concern healthcare providers, healthcare professionals, and manufacturers of medical devices, among others.

*Proposed laws*

The AI Act proposal may impact the cybersecurity of AI medical devices in four ways. The Act, in its current wording, contains provisions prohibiting certain AI systems and practices, proposes a risk-based mechanism for governing those AI systems that pose a high risk for individuals or society, stipulates fines for providers' non-compliance with the Act, and establishes an EU body responsible for the harmonised application of the Act amongst the Member States.[79] The AI Act proposal explicitly includes medical devices in the scope of its application in Article 6 and Annex II. Based on the current wording of the AI Act proposal, the definitions of AI and risk classification provided therein imply that any medical device software could fall within the scope of the AI Act and be considered a high-risk AI system since most medical device software needs a conformity assessment by a notified body.[80] Consequently, this could imply parallel application of the two pieces of legislation and pose additional challenges for those implementing them (further elaborated in the following section). What concerns the cybersecurity of AI medical devices, Article 13(1) of the proposal requires that the design and development of high-risk AI systems are done in a way that ensures their transparent operation so the users can interpret the system's output and use it appropriately. They shall be designed and developed to achieve an appropriate level of accuracy, robustness and cybersecurity and perform consistently throughout their lifecycle (Article 15(1)). In the instructions for use (Article 15(2-3)), providers shall specify the level against which cybersecurity of the system has been tested and validated, which can be expected, and any known and foreseeable circumstances that may impact that level of cybersecurity. Article 15(4) of the proposal requires that the technical solutions aimed at ensuring the cybersecurity of high-risk AI systems are appropriate to the relevant circumstances and the risks. To this end, high-risk AI systems certified according to the Cybersecurity Act shall be presumed to comply with the cybersecurity requirements set out in the proposal (AI Act, Article 42).

---

[71] ibid., req 14.2.(d).

[72] ibid, req 14.5.

[73] ibid, req 17.1.

[74] idem.

[75] ibid, req 17.2.

[76] ibid, req 17.4.

[77] ibid, see reqs 23.1.(g), 23.2.(m), 23.4.(ab).

[78] GDPR, arts 5 and 32.

[79] In this respect, see also Maximilian Gartner, 'Why the prohibition of certain persuasive technologies in the European proposal for an artificial intelligence act is not a surprise' (*KU Leuven CiTiP Blog*, 29 April 2021) < https://www.law.kuleuven.be/citip/blog/why-the-prohibition-of-certain-persuasive-technologies-in-the-european-proposal-for-an-artificial-intelligence-act-is-not-a-surprise/> accessed 11 August 2022.

The NIS 2 Directive proposal is set to reform the currently applicable NIS Directive due to the varying level of harmonisation of the NIS Directive among the EU Member States. What concerns cybersecurity of AI medical devices and in comparison with the applicable NIS Directive, the proposal removes the Member States' requirement to identify OES and Digital Service Providers (DSP) in their territories. The proposal also replaces OES and DSPs with new categories: 'essential' and 'important entities' (enlisted in Annexes I and II of the proposal). They are ordered per sectors and sub-sectors (for example, 'health' and 'manufacturing' sectors; 'manufacture of medical devices and in-vitro medical diagnostic medical devices' sub-sectors). Every sub-sector contains a list of 'types of entities'. The NIS 2 Directive proposal broadens its scope of application. For instance, healthcare providers are now considered 'essential entities' (Annex I). In addition, the NIS 2 Directive proposal adds new types of entities relevant to the healthcare sector.[81] Similarly to the NIS Directive, the NIS 2 Directive proposal mandates the Member States to establish security measures for the entities under its scope. Chapter IV of the proposal contains the obligations on cybersecurity and risk management and reporting. Article 18 of the proposal on cybersecurity risk management measures implies that essential and important entities shall 'take appropriate and proportionate technical and organisational measures to manage the risks posed to the security of network and information system'. As examples of measures, the article includes, amongst others, incident handling (prevention, detection, and response to incidents) and measures to ensure supply chain security and vulnerability handling and disclosure. Article 20 of the proposal on reporting obligations introduces a two-step procedure to report significant security breaches, which could also be reported to the recipients of their services. Article 21 of the proposal concerns cybersecurity certification schemes. Enforcement and supervision of essential and important entities are delegated to competent authorities. Competent authorities shall supervise them and ensure their compliance with the security and incident notification requirements. An ex-ante supervisory regime is in place for essential entities and an ex-post one for important entities.

---

[80] MedTech Europe, 'Proposal for an Artificial Intelligence Act (COM/2021/206) - MedTech Europe response to the open public consultation' (2021) <https://www.medtecheurope.org/wp-content/uploads/2021/08/medtech-europe-response-to-the-open-public-consultation-on-the-proposal-for-an-artificial-intelligence-act-6-august-2021-1.pdf> accessed 25 October 2022.

[81] For further analysis, see Biasin and Kamenjašević (n 4).

## Current Challenges in the Regulatory Framework

### The AI Act proposal and its interaction with MDR: cybersecurity and safety requirements

Both the MDR and the AI Act proposal requirements are relevant from the cybersecurity perspective. To demonstrate this, we propose the example of an anaesthesia device.[82] In this example, an unauthorised user with physical access to an anaesthesia device, thanks to social engineering techniques, guesses the weak password of the device and manipulates its settings so that the machine may supply a wrong anaesthetic concentration.

Following this case, both the MDR and the AI Act proposal could be relevant. It could be a serious incident under the MDR, as the wrong anaesthetic concentration might (even indirectly) lead to the serious deterioration of a patient's state of health. For the same reason (i.e., the serious deterioration of a patient's health), the event could also be relevant under the AI Act proposal.

In some circumstances, the AI Act proposal requires providers of high-risk AI systems to report serious incidents to the relevant authority.[83] Similarly, the MDR also requires the reporting of serious incidents to the medical devices competent authorities.[84] A consequent first challenge concerns the overlapping of cybersecurity requirements between the MDR and the AI Act proposal, in particular on the matter of serious incident notification. The overlapping of the MDR's and AI Act proposal's requirements is not necessarily a problem *per se*. Nevertheless, it remains essential to understand the interaction between the safety requirements of the AI Act with the MDR and whether the MDR should be considered as a *lex specialis* to the AI Act or not.

The AI Act proposal and the MDR present similarities but also divergences concerning serious incidents. Elsewhere, we demonstrated that the definition and some other aspects concerning serious incident reporting might differ.[85] For example, the regulated entities are different. In the first case, the AI Act proposal covers high-risk systems providers, while the MDR concerns medical device manufacturers. These may or may not overlap. Manufacturers of medical devices can also be providers if their product is software or hardware with software, but they may also license or buy these parts separately, which would not make them providers in light of the AI Act. As another crucial difference, the AI Act proposal relies on the 'breach of obligations to protect individuals fundamental rights' as a condition for reporting serious incidents. The MDR is also about the individuals' fundamental right to access to health care and safety, but it does not subordinate the respect of fundamental rights as a condition for the notification of serious incidents.[86]

---

[82] see also MDCG guidance (n 6).

[83] see AI Act proposal, art 62, according to which serious incidents are any 'incident that directly or indirectly led, might have led, or might lead to (…) (a) the death of a person or serious damage to a person's health, to property or the environment, (b) a serious and irreversible disruption of the management and operation of critical infrastructure').

[84] see MDR, art 87, according to which serious incidents are 'any incident that directly or indirectly led, might have led or might lead to (…) the death of the patient user or other person, the temporary or permanent serious deterioration of a patient's, user's or other person's state of health, a serious public threat'.

[85] Biasin and Kamenjašević (n4 ).

[86] As part of the risk management obligations, see MDR art 10(2); see also Annex I, req 18.8.

In its Explanatory Memorandum, the AI Act proposal states that the proposal itself 'will be integrated into the existing sectoral safety legislation to ensure consistency'.[87] This formulation in the AI Act proposal text lacks specificity: it is not clear how could this 'integration' happen between the different pieces of legislation.

In conclusion, a possible challenge for the AI Act proposal could be understanding if and how the safety requirements will be integrated into medical devices sector-specific legislation. The references provided in the actual text of the AI Act proposal seem not articulated enough to help understand how safety requirements are going to interact with the MDR. If these integration aspects are not timely addressed, the lack of coordinated frameworks could lead to regulatory uncertainty.[88]

### The NIS 2 Directive proposal and its interaction with the MDR: incident notification requirements

The NIS 2 Directive proposal and the MDR also include relevant requirements from the cybersecurity perspective. In fact, serious incidents under the MDR may qualify as also as incidents under the NIS 2 Directive proposal. We offer an example to demonstrate it, by making the case of an implantable sensor used to monitor pulmonary artery pressures in heart failure patients.[89] In our example, an adversary modifies through data poisoning or creates patient data in transit to or from the external electronics unit, causing misdiagnosis affecting patient care. As a consequence of safety harm, the physician could fail to provide the treatment based on incorrect low pulmonary artery pressure readings. This situation could lead to the worsening of the patient's heart failure condition.[90] In this case, worsening a patient's heart failure condition could be seen as a deterioration of their health caused by the event initiated by the attacker (relevant under the MDR's serious incident requirements). The misdiagnosis of patient care could also be considered as impacting the provision of healthcare services, thus relevant from the NIS 2 Directive proposal perspective.

As illustrated above, the MDR sets obligations concerning serious incident reporting. The NIS 2 Directive proposal requires that any incident having a significant impact on the provision of essential or important entities' services shall be notified to the national competent authority or CSIRT.[91]

The NIS 2 Directive proposal entails similar challenges in terms of regulatory convergence with the MDR.[92] The situation described above could imply the application of both the MDR and NIS 2 Directive requirements. Such a parallel application of requirements is where the challenges might start. In the past, the European Commission assessed the possible "synergies" of the NIS Directive with sector-specific regulation, including in the medical devices field concerning incident notification reporting.[93]

---

[87] AI Act proposal, Explanatory Memorandum, 4.

[88] ibid.

[89] The example is taken from the MDCG guidance (n 6).

[90] ibid.

[91] For a definition of essential and important entities, see NIS 2 Directive proposal art 4(25)-(26). Concerning incident notfication, it is worth noting that the current formulation of the proposal includes also references to 'any significant cyber threat that those entities identify that could have potentially resulted in a significant incident' (NIS 2 Directive proposal, art 20(2)2)

[92] Medical device manufacturers are subject to both obligations. For the NIS 2 Directive proposal, it is worth to note that – being it a Directive – its requirements will have to be set by national legislation.

[93] NIS Cooperation Group, 'Synergies in Cybersecurity Incident Notification Reporting' (2020).

The actual text of the NIS 2 Directive proposal suggests that where provisions of sector-specific acts of Union law require medical device manufacturers to notify incidents or significant cyber threats, the provisions of the NIS 2 Directive should not apply *if* those requirements are "at least equivalent".[94] The provision may turn problematic, for example, for manufacturers assessing which piece of legislation should apply. In this regard, it is questionable whether the MDR's safety requirements should be considered as *at least equivalent* to the NIS2 Directive's ones concerning incident notification. In fact, safety incidents do not always equate to security incidents.[95]

We could take the example of a warming device for premature babies.[96] The MDCG provides the following example: an unauthorised user with physical access to the device guesses the weak password for the service account and exports therapy and patient data via the USB interface. In the view of the MDCG, there could be security harm (i.e., the unauthorised access) which could not result in safety harm (e.g., the serious deterioration of a patient's health) in terms of MDR's serious incident notification rules.[97] In other words, medical device manufacturers would have to notify the incident about the unauthorised access to the NIS 2 Directive proposal competent authority and not the national relevant authority under the MDR.

Given that security incidents do not always equate to safety incidents, one may wonder which direction should be taken in interpreting the MDR and the NIS 2 Directive proposal's rules. We envisage at least three possible approaches for the legislator. A first, simplistic approach would mean considering the MDR as "at least equivalent" to the NIS 2 Directive proposal. In this case, the NIS 2 Directive requirements on incident notification would not apply, while the MDR would prevail. A second approach could, on the contrary, suggest the parallel application of both legal acts and their requirements. Such a parallel application would require manufacturers to take into account the different requirements and apply them both. A third approach could consider the MDR as a *lex specialis*, leaving the NIS 2 Directive as possible general legislation to cover those hypotheses that are not strictly covered by the MDR but are still relevant to the NIS 2 Directive proposal.

The solutions we propose come with their own advantages and disadvantages.[98] It remains essential that the legislator takes a stance and resolves this interpretative issue to ensure regulatory certainty for all stakeholders in healthcare.

### The evolving term 'critical infrastructures': serious incidents for AI systems in the healthcare sector

A further challenge within the AI Act proposal stems from the 'serious incident' definition. Serious incidents are any incident that directly or indirectly led, might have led, or might lead to a 'serious and irreversible disruption of the management and operation of critical infrastructure' (Article 3(44) of the AI Act proposal, second part).[99]

---

[94] NIS2 Directive proposal, art 2(6).

[95] Safety will matter when cybersecurity causes injury or damage, see Ross Anderson, *Security Engineering: A Guide to Building Dependable Distributed Systems (John Wiley & Sons 2020)*, 1044 – 1045. But many situations will not involve safety, especially when the cyberattack considers privacy.

[96] MDCG guidance (n 6).

[97] ibid.

[98] For a more in-depth analysis on the proposed solutions, advantages and disadvantages, see also Biasin and Kamenjašević (n 4).

The identification of critical infrastructures, however, is delegated to the EU Member States, following the principle of subsidiarity.[100] In simple terms, the division of competencies between the EU and its Member States entails that the identification of a hospital, healthcare entity or health process as a critical infrastructure shall ultimately depend on every national approach adopted in critical infrastructure regulation.[101] Until recently, the EU Member States have adopted various approaches to identify their own critical infrastructures. Usually, the trends followed by the Member States see the definition of critical infrastructure based on their defence strategies, national emergency management and long-term national traditions.[102]

Studies concerning the identification of critical infrastructures in the healthcare sector have shown how heterogeneous the situation could be. For instance, France includes the healthcare sector within the scope of critical infrastructure protection legislation, and the Netherlands has embraced a process-oriented approach that does not include healthcare assets.[103] This conceptual issue may entail some tangible consequences. We can use the following example: a cyberattack carried out through social engineering techniques or extraction of data on a healthcare provider. Some Member States could identify the event as an attack on critical infrastructure, and others may not because healthcare is not considered a critical infrastructure sector. The first scenario would unlock the condition to possibly consider the event as a serious incident and thus activate the obligation of its reporting pursuant to the AI Act proposal. The second case would likely not entail it so.[104]

AI systems are used in healthcare critical infrastructures. If disrupted by a cyberattack, the provision of healthcare services to individuals may affect the continuity of their services and, ultimately, the quality of healthcare systems. The Member States not considering healthcare within the sectors or processes of critical infrastructures may lower the level of protection of individuals, if compared to the Member States who do it, from the perspective of the AI Act incident notification rules. In the ultimate analysis, this kind of heterogeneous level of individuals' protection across the EU may be seen as a possible regulatory challenge, which could even lead to fragmentation risks in the EU internal market.[105]

---

[99] AI Act proposal, art 3(44).

[100] From a conceptual perspective, the notion 'critical infrastructures' has been defined as an evolving term. See Dimitra Markopoulou and Vagelis Papakonstantinou, 'The Regulatory Framework for the Protection of Critical Infrastructures against Cyberthreats: Identifying Shortcomings and Addressing Future Challenges: The Case of the Health Sector in Particular' (2021) 41 Computer Law & Security Review 105502. For the purpose of this book chapter, we refer to the definition of Rinaldi, Perenboom & Kelly: Infrastructures are considered 'critical' when their disruption could have an impact on the functioning of the society (in terms of economy, security and people's wellbeing. Steven M Rinaldi, James P Peerenboom, and Terrence K Kelly, 'Identifying, Understanding, and Analyzing Critical Infrastructure Interdependencies' (2001) IEEE Control Systems Magazine 11. For an overview of the principles guiding the implementation of critical infrastructure protection, see also, Commission, 'Communication from the Commission on a European Programme for Critical Infrastructure Protection COM (2006) 786 final' (2006), 3.

[101] There exist a EU level definition of 'European Critical Infrastructure', which is determined in the European Critical Infrastructure Directive (ECI Directive) – see Council Directive 2008/11/EC of 8 December 2008 on the identification and designation of European critical infrastructures and the assessment of the need to improve their protection [2008] OJ L345/75 (ECI Directive). Nevertheless, national critical infrastructures do not belong to the ECI Directive scope and the Member States have autonomy in determining their approaches towards them.

[102] Elisabetta Biasin and others, SAFECARE D3.9, 48.

[103] ibid. See also France, Ordonnance n° 2004-1374 du 20 décembre 2004 (Code de la Défense).

[104] Ultimately, this would also depend whether lett (a) of art 33(44) of the AI act proposal is met.

As a possible step forward, these uneven impacts could be mitigated by shifting the focus of Article 33 AI Act proposal from 'critical infrastructures' to 'critical entities'.[106] In fact, in recent years, the many issues surrounding the European legislation concerning the protection of critical infrastructures led the European legislator to revise the existing legislation concerning the European Critical Infrastructures to issue a new proposed Directive on 'critical entities'.[107] While, as outlined above, critical infrastructures do not meet the same criteria in terms of sectors at the national level, the CER Directive proposal could help fill this gap because it includes the health sector within its scope.[108] In other words, the CER Directive, once approved, will enlist the sectors under which the Member States will have to identify the critical entities where healthcare is part of them.[109] Therefore, the risk of not considering healthcare as part of the scope of protection would be mitigated by the directive's text. In conclusion, considering 'critical entities' instead of 'critical infrastructures' could minimise the problem of uneven consideration of the healthcare sector for serious incidents concerning AI systems in the EU.

## Conclusion

The purpose of this book chapter was to draw up and comment on the essential aspects concerning cybersecurity of AI as or in medical devices from integrated security and legal perspective. In the first part of the book chapter, we explained how cybersecurity threats might be in continuous evolution. To underscore the problem's relevance, we selected three examples: dataset poisoning, social engineering and data or source code extraction. In the following section, we integrated the security parts with the description of the legal aspects surrounding medical device cybersecurity. We analyzed the MDR, the NIS Directive, and the Cybersecurity Act. Regulatory challenges arise as the legal framework concerning AI and cybersecurity evolves. Therefore, we illustrated some of the core challenges of the two new proposed pieces of legislation: the NIS 2 Directive and the AI Act proposals. The AI and cybersecurity regulations are recent matters of regulation, and future research will have to closely monitor their future regulatory challenges.

---

[105] Biasin and Kamenjašević (n 4).

[106] Similarly to the NIS Directive, the CER Directive proposes Member States to identify critical entities based on common criteria for national risk assessments. For the definition of critical entities, see Commission, 'Proposal for a Directive of the European Parliament and of the Council on the resilience of critical entities' COM/2020/829 final (hereinafter CER Directive proposal), art 2(1).

[107] On these issues, see Elisabetta Biasin, 'Healthcare Critical Infrastructures Protection and Cybersecurity in the EU: Regulatory Challenges and Opportunities' (Proceedings of the 1st European Cluster for Securing Critical Infrastructures (ECSCI) Virtual Workshop, June 2020).

[108] see Annex of the CER Directive proposal. There could be counterarguments, however, against this proposal, which for reasons of space it is not possible to analyse with due detail in this chapter. Further discussions about this proposal may question the more expanded scope of critical entities' sectors against critical infrastructures ones following the ECI Directive. We limit ourselves to observe that this argument may be counter-replied by observing that the scope of critical infrastructure protection can be nevertheless expanded by the Member States at the national level. Comparison and illustration of the differences between critical entities and critical infrastructures would also require more space; to this purpose we refer to the historical and conceptual analysis by Markopoulou and Papakonstantinous (2020).

[109] The identification of critical entities would have to follow specific criteria and procedures, see CER Directive proposal, art 5.

# Bibliography

Abdelfattah and others, 'Adversarial Attacks on Camera-LiDAR Models for 3D Car Detection', 2021 IEEE/RSJ International Conference on Intelligent Robots and Systems (IROS) (IEEE 2021) <https://ieeexplore.ieee.org/document/9636638/> accessed   25 October 2022

Abroshan and others, 'COVID-19 and Phishing: Effects of Human Emotions, Behavior, and Demographics on the Success of Phishing Attempts During the Pandemic' (2021) 9 IEEE Access 121916

Al-Hamar J and others, 'Enterprise Credential Spear-Phishing Attack Detection' (2021) 94 Computers & Electrical Engineering 107363

Alemzadeh H and others, 'Targeted Attacks on Teleoperated Surgical Robots: Dynamic Model-Based Detection and Mitigation' (2016) Proceedings - 46th Annual IEEE/IFIP International Conference on Dependable Systems and Networks, DSN 2016 395

Anderson R, *Security Engineering: A Guide to Building Dependable Distributed Systems* (John Wiley & Sons 2020)

Arif Z H and others, 'Comprehensive Review of Machine Learning (ML) in Image Defogging: Taxonomy of Concepts, Scenes, Feature Extraction, and Classification Techniques' (2022) 16 IET Image Processing 289

Biasin E and Kamenjašević E, 'Cybersecurity of medical devices: new challenges arising from the AI Act and NIS 2 Directive proposals' [2022] International Cybersecurity Law Review, 163, https://doi.org/10.1365/s43439-022-00054-x.

Biasin E and Kamenjasevic E, 'Cybersecurity of Medical Devices: Regulatory Challenges in the European Union' in Carmel Shachar and others (eds), The Future of Medical Device Regulation: Innovation and Protection (Cambridge University Press 2022) <https://www.cambridge.org/core/books/future-of-medical-device-regulation/cybersecurity-of-medical-devices/AC01289C2DB05E44D0D98A9E66666562>

Biasin E, 'Healthcare Critical Infrastructures Protection and Cybersecurity in the EU: Regulatory Challenges and Opportunities' (Proceedings of the 1st European Cluster for Securing Critical Infrastructures (ECSCI) Virtual Workshop, June 2020)

Bullée J and Junger M, 'Social Engineering' in Thomas J Holt and Adam M Bossler (eds), The Palgrave Handbook of International Cybercrime and Cyberdeviance (Springer International Publishing 2020) <http://link.springer.com/10.1007/978-3-319-78440-3_38> accessed   25 October 2022

Braga A and Logan RK, 'AI and the Singularity: A Fallacy or a Great Opportunity?' (2019) 10 Information

Bygrave L A, 'The "Strasbourg Effect" on Data Protection in Light of the "Brussels Effect": Logic, Mechanics and Prospects' (2021) 40 Computer Law & Security Review 105460

[PRE-PRINT/PEER REVIEWED]   4.0 –   14/12/2022

Castryck W and Decru T, 'AN EFFICIENT KEY RECOVERY ATTACK ON SIDH (PRELIMINARY VERSION)' <https://eprint.iacr.org/2022/975.pdf> accessed     25 October 2022

Celi S, 'The Quantum Solace and Spectre' (The Cloudflare Blog, 21 February 2022) <http://blog.cloudflare.com/quantum-solace-and-spectre/> accessed     25 October 2022

Christen M, Gordijn B and Loi M (eds), The Ethics of Cybersecurity, vol 21 (Springer International Publishing 2020) <http://link.springer.com/10.1007/978-3-030-29053-5> , chapter 7.

Commission, 'Communication from the Commission on a European Programme for Critical Infrastructure Protection COM (2006) 786 final' (2006)

Commission, 'Proposal for a Regulation of the European Parliament and of the Council Laying Down Harmonised Rules on Artificial Intelligence (Artificial Intelligence Act) and Amending Certain Union Legislative Acts' COM/2021/206 final

Commission 'Proposal for a Directive of the European Parliament and of the Council on measures for a high common level of cybersecurity across the Union, repealing Directive (EU) 2016/1148' COM(2020) 823 final

Commission, 'Proposal for a Directive of the European Parliament and of the Council on the resilience of critical entities' COM/2020/829 final

Council Directive 2008/11/EC of 8 December 2008 on the identification and designation of European critical infrastructures and the assessment of the need to improve their protection [2008] OJ L345/75 (ECI Directive)

Directive (EU) 2016/1148 of the European Parliament and of the Council of 6 July 2016, concerning measures for a high common level of security of network and information systems across the Union [2016] OJ L 194/1

Pai D B, 'Mapping the Genealogy of Medical Device Predicates in the United States' (2021) 16 PLOS ONE e0258153.

ECOMMERCE EUROPE, 'Position Paper on the Artificial Intelligence Act' (2022) <https://ecommerce-europe.eu/wp-content/uploads/2022/02/ECOM-AI-Position-Paper-01022022.pdf> accessed    25 October 2022

Fuertes W and others, 'Impact of Social Engineering Attacks: A Literature Review' in Álvaro Rocha, Carlos Hernan Fajardo-Toro and José María Riola Rodríguez (eds), Developments and Advances in Defense and Security (Springer Singapore 2022)

Gartner M,  'Why the prohibition of certain persuasive technologies in the European proposal for an artificial intelligence act is not a surprise' (KU Leuven CiTiP Blog, 29 April 2021) < https://www.law.kuleuven.be/citip/blog/why-the-prohibition-of-certain-persuasive-technologies-in-the-european-proposal-for-an-artificial-intelligence-act-is-not-a-surprise/> accessed    25 October 2022

[PRE-PRINT/PEER REVIEWED] 4.0 – 14/12/2022

Gong X and others, 'Model Extraction Attacks and Defenses on Cloud-Based Machine Learning Models' (2020) 58 IEEE Communications Magazine 83

Jagielski M and others, 'Manipulating Machine Learning: Poisoning Attacks and Countermeasures for Regression Learning' (2018) 2018-May Proceedings - IEEE Symposium on Security and Privacy 19

Kaviani S and others, 'Adversarial Attacks and Defences on AI in Medical Imaging Informatics: A Survey' (2022) Expert Systems With Applications 116815 <https://doi.org/10.1016/j.eswa.2022.116815>

Kobara K, 'Cyber Physical Security for Industrial Control Systems and IoT' (2016) E99D IEICE Transactions on Information and Systems 787

Khoo C and others, 'Installing Fear: A Canadian Legal and Policy Analysis of Using, Developing, and Selling Smartphone Spyware and Stalkerware Applications' (2019) < https://citizenlab.ca/2019/06/installing-fear-a-canadian-legal-and-policy-analysis-of-using-developing-and-selling-smartphone-spyware-and-stalkerware-applications/> accessed 25 October 2022

Kuzlu M and others, 'Role of Artificial Intelligence in the Internet of Things (IoT) Cybersecurity' (2021) 1 Discover Internet of Things <https://doi.org/10.1007/s43926-020-00001-4>

Hutchings A, Pastrana S and Clayton R, 'Displacing Big Data: How Criminals Cheat the System', The Human Factor of Cybercrime (Routledge 2017)

International Medical Device Regulators Forum, '" Software as a Medical Device": Possible Framework for Risk Categorization and Corresponding Considerations' (2014

Gal MS and Aviv O, 'The Competitive Effects of the GDPR' (2020) 16 Journal of Competition Law & Economics 349

Galtan S, 'Healthcare giant Magellan Health hit by ransomware attack' (Bleeping Computer, 12 May 2020) <https://www.bleepingcomputer.com/news/security/healthcare-giant-magellan-health-hit-by-ransomware-attack/>, accessed 25 October 2022

HIPAA Journal, 'Healthcare Data Breaches Due to Phishing' (HIPAA Journal, nd) <https://www.hipaajournal.com/healthcare-data-breaches-due-to-phishing/>, accessed 25 October 2022

Levy-Loboda T and others, 'Personalized Insulin Dose Manipulation Attack and Its Detection Using Interval-Based Temporal Patterns and Machine Learning Algorithms' (2022) 132 Journal of Biomedical Informatics 104129

Ludvigsen K and Nagaraja S, 'Dissecting Liabilities in Adversarial Surgical Robot Failures: A National (Danish) and EU Law Perspective' (2022) 44 Computer Law and Security Review <https://doi.org/10.1016/j.clsr.2022.105656>

Ludvigsen K, Nagaraja S and Daly A, 'When Is Software a Medical Device? Understanding and Determining the "Intention" and Requirements for Software as a Medical Device in European Union Law' [2021] European Journal of Risk Regulation <https://doi.org/10.1016/j.clsr.2022.105656>

Markopoulou D and Papakonstantinou V, 'The Regulatory Framework for the Protection of Critical Infrastructures against Cyberthreats: Identifying Shortcomings and Addressing Future Challenges: The Case of the Health Sector in Particular' (2021) 41 Computer Law & Security Review 105502

Medical Device Coordination Group, 'MDCG 2019-16 Guidance on Cybersecurity for Medical Devices, December 2019' (2020)

MedTech Europe, 'Proposal for an Artificial Intelligence Act (COM/2021/206) - MedTech Europe response to the open public consultation' (*MedTech Europe,* 2021) <https://www.medtecheurope.org/wp-content/uploads/2021/08/medtech-europe-response-to-the-open-public-consultation-on-the-proposal-for-an-artificial-intelligence-act-6-august-2021-1.pdf> accessed 25 October 2022

NIS Cooperation Group, 'Reference document on security measures for Operators of Essential Services' (2018)

NIS Cooperation Group, 'Synergies in Cybersecurity Incident Notification Reporting' (2020)

Pai D B, 'Mapping the Genealogy of Medical Device Predicates in the United States' (2021) 16 PLOS ONE e0258153

Pan X and others, 'How You Act Tells a Lot: Privacy-Leaking Attack on Deep Reinforcement Learning' (2019) 1 Proceedings of the International Joint Conference on Autonomous Agents and Multiagent Systems, AAMAS 368

Papernot N and others, 'Practical Black-Box Attacks against Machine Learning' [2017] ASIA CCS 2017 - Proceedings of the 2017 ACM Asia Conference on Computer and Communications Security 506

Peukert C and others, 'Regulatory Spillovers and Data Governance: Evidence from the GDPR' (2022) Marketing Science <https://pubsonline.informs.org/doi/pdf/10.1287/mksc.2021.1339> accessed 25 October 2022

Regulation (EU) 2016/679 of the European Parliament and of the Council of 27 April 2016 on the protection of natural persons with regard to the processing of personal data and on the free movement of such data, and repealing Directive 95/46/EC (General Data Protection Regulation) [2016] OJ L 1191

Regulation (EU) 2017/745 of the European Parliament and of the Council of 5 April 2017, on medical devices, amending Directive 2001/83/EC, Regulation (EC) No 178/2002 and Regulation (EC) No 1223/2009 and repealing Council Directives 90/385/EEC and 93/42/EEC [2017] OJ L 117/1

Regulation (EU) 2019/881 of the European Parliament and of the Council of 17 April 2019 on ENISA (the European Union Agency for Cybersecurity) and on information and communications technology cybersecurity certification and repealing Regulation (EU) No 526/2013 (Cybersecurity Act) [2019] OJ L 151/15

Rinaldi S, Peerenboom J and Kelly T, 'Identifying, Understanding, and Analyzing Critical Infrastructure Interdependencies' (2001) 21 IEEE Control Systems Magazine 11

Savadjiev P and others, 'Demystification of AI-Driven Medical Image Interpretation: Past, Present and Future' (2019) 29 European Radiology 1616, 1616–1618

Shimron E and others, 'Subtle Data Crimes: Naively Training Machine Learning Algorithms Could Lead to Overly-Optimistic Results' (2021) <http://arxiv.org/abs/2109.08237> accessed 25 October 2022

Swain S and Niyogi R, 'SmartMedicist: A Context-Aware System for Recommending an Alternative Medicine' (2018) 14 International Journal of Pervasive Computing and Communications 147

Tomsett R and others, 'Why the Failure? How Adversarial Examples Can Provide Insights for Interpretable Machine Learning' [2018] 2018 21st International Conference on Information Fusion, FUSION 2018 838

Tschider CA, 'Deus Ex Machina: Regulating Cybersecurity and Artificial Intelligence for Patients of the Future' (2018) 5 Savannah L Rev 177

US Food and Drug Administration 'How to Determine if Your Product is a Medical Device' (*FDA.gov*, December 2019) <https://www.fda.gov/medical-devices/classify-your-medical-device/how-determine-if-your-product-medical-device> accessed 25 October 2022

Xu H and others, 'Transferable Environment Poisoning: Training-Time Attack on Reinforcement Learning' (2021) 3 Proceedings of the International Joint Conference on Autonomous Agents and Multiagent Systems, AAMAS 1386.

Yampolskiy RV, 'AI Is the Future of Cybersecurity, for Better and for Worse' [2017] Harvard Business Review <https://hbr.org/2017/05/ai-is-the-future-of-cybersecurity-for-better-and-for-worse> accessed 25 October 2022

Zhou D and others, 'Removing Adversarial Noise in Class Activation Feature Space', 2021 IEEE/CVF International Conference on Computer Vision (ICCV) (IEEE 2021) <https://ieeexplore.ieee.org/document/9710484/> accessed 25 October 2022